\documentclass[12pt]{article}
\usepackage{amsmath}
\usepackage{amsthm}
\usepackage{amssymb}
\usepackage[english]{babel}
\usepackage[dvips]{graphicx}

\newtheorem{theorem}{Theorem}
\newcommand{\g}{\gamma}
\newcommand{\R}{\mathbf{R}}
\newcommand{\T}{\mathbf{T}}

\begin{document}

\begin{flushright}
\footnotesize

Soviet Math. Dokl.

Vol. 28 (1983), No. 3, pp. 802-805

AMS 0197-6788/84

\end{flushright}

\normalsize

\begin{center}
{\bf TOPOLOGICAL ANALYSIS

OF CLASSICAL INTEGRABLE SYSTEMS

IN THE DYNAMICS OF THE RIGID BODY}

\end{center}

\noindent UDC 513.83:531.38

\begin{center}
{\bf M. P. KHARLAMOV}

\end{center}

The equations of motion of a rigid body around a fixed point
\begin{equation}\label{eq01}
\dot \nu = \nu \times \omega, \quad {\bf A}\dot \omega+\omega \times
{\bf A}\omega={\bf e}\times \nu
\end{equation}
(in which $\bf A$ is the inertia tensor, $\omega$ is the angular
velocity, $\nu$ is the unit vertical vector, and $\bf e$ is the
radius-vector of the center of mass, and all the quantities are
related to the moving coordinate axes) define a dynamical system on
the phase space $M = S^2\times {\bf R}^3$ and possess the first
integrals
\begin{equation}\label{eq02}
H=\frac{1}{2}{\bf A}\omega \cdot \omega-{\bf e}\cdot \nu, \quad
G={\bf A}\omega \cdot \nu.
\end{equation}

It is known that equations (\ref{eq01}) can be integrated by
quadratures if, in addition to (\ref{eq02}), there exists another
(possibly, partial) integral (see \cite{Gol}--\cite{Koz})
\begin{equation}\label{eq03}
K=K(\nu,\omega).
\end{equation}

Of special interest are the general integrability cases: the
solutions of Lagrange, Euler, Kovalevskaya, and Goryachev-Chaplygin.
The first two can be included in Smale's scheme for studying the
phase topology of natural systems \cite{Sma}; their analysis is
carried out in \cite{Iac}. The topology of the Euler case was
considered also in \cite{Kh76} and \cite{Lac}. The present article
is devoted to the investigation of the integral manifolds
$$
J_n = \{z \in M: H(z) = h, K(z) = k, G(z) = g\}, \quad n = (h, k, g)
\in {\bf R}^3,
$$
in the cases of Kovalevskaya and Goryachev-Chaplygin.

We modify Smale's program for topological analysis of mechanical
systems  \cite{Abr} to suit the specific problem under
consideration. Thus, in order to describe the phase topology of
system (\ref{eq01}) with integrals (\ref{eq02}) and (\ref{eq03}) it
is necessary to construct the bifurcation set $\Sigma$ of the map
${H \times K \times G: M \to {\bf R}^3}$, to determine the
topological type of $J_n$ for all admissible values $n \in {\bf
R}^3$ and the structure of the phase flow on $J_n$, and to indicate
all types of topological modifications that $J_n$ undergoes when the
point $n$ crosses $\Sigma$. In the case $n \notin \Sigma$ the
manifold $J_n$ is a union of a finite number of two-dimensional tori
carrying quasi-periodic trajectories \cite{Arn}, \cite{Koz}. It
remains only to determine the number of components of $J_n$. A more
difficult problem is to describe the structure of the critical
integral surfaces. The results formulated below rest upon an
analysis of the projection of $J_n$ on the Poisson sphere
\cite{Kh82}.

1. In the Kovalevskaya case, introducing suitable measures, we
represent the integrals (\ref{eq02}) and (\ref{eq03}) in the form
$$
\begin{array}{l}
H = \omega_1^2+\omega_2^2+\frac{1}{2}\omega_3^2-\nu_1, \quad
G=\omega_1 \nu_1+\omega_2 \nu_2+\frac{1}{2}\omega_3 \nu_3,\\
K=(\omega_1^2-\omega_2^2+\nu_1)^2+(2\omega_1 \omega_2 +\nu_2)^2.
\end{array}
$$

\begin{theorem}
The sections $\Sigma_g$ of the bifurcation set $\Sigma$ by planes $g
= \mathrm{const}$ are defined by the equalities
$$
\begin{array}{l}
\Sigma_0=\{k=0,h \geqslant 0\}\cup \{k=1,h \geqslant -1\}\cup
\{k=h^2,h \geqslant -1\}\cup \{k=h^2+1,h
\geqslant 0\},\\[2mm]
\displaystyle{\Sigma_g=\{k=(h-2g)^2,g^2-1 \leqslant h \leqslant
\frac{1}{4g^2}+2g^2\}\cup\{k=0,h\geqslant 2g^2\}}\\[3mm]
\qquad \displaystyle{\cup \{h=\tau+\frac{g^2}{\tau^2},
k=1-\frac{2g^2}{\tau}+\frac{g^4}{\tau^4}: -1 \leqslant \tau <0,
0<\tau<+\infty; k>0\}.}
\end{array}
$$
The surface $\Sigma$ divides ${\bf R}^3$ into $6$ open components.
The corresponding integral manifolds are $\varnothing$, ${\bf T}^2$,
$2{\bf T}^2$ {\rm (}for three components{\rm )}, and $4{\bf T}^2$.
\end{theorem}

In Figures 1 and 2 the sets $\Sigma_g$ are indicated in the richest
cases $0 < g^2 < \frac{1}{2}$ and ${\frac{4}{3\sqrt{3}} < g^2 < 1}$.
The numerals denote the number of components of $J_n$.

To describe the character of bifurcations we introduce the following
notation: $V$ is the wedge of two circles, or a "figure-eight"; $W$
is a set homeomorphic to the intersection of a two-dimensional
sphere with a pair of planes passing through its center (i.e., two
circles that intersect at two points); $P$ is the skew product of a
circle by a "figure-eight"\, (obtained from $[0,1] \times V$
identifying $\{0\} \times V$ and $\{1\} \times V$ by means of a map
homotopic to the central symmetry of the "figure-eight"; see
Figure~3); $Q = W \times S^1$, and $R = V \times S^1$.

\begin{figure}[h]\label{fig1}
\begin{center}
\includegraphics[width=160mm,keepaspectratio]{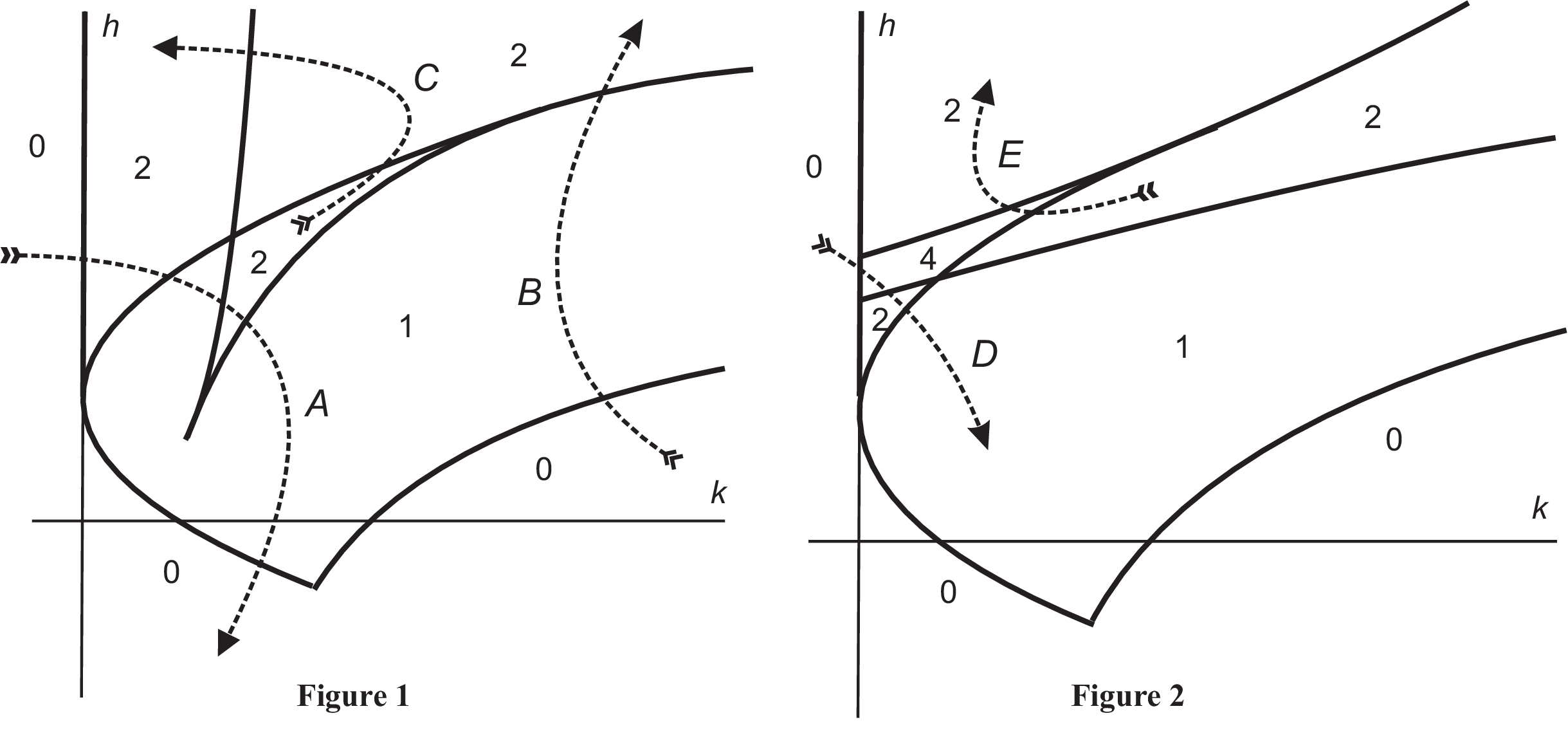}
\end{center}
\end{figure}

Let $\g$ be the boundary of the $\varepsilon$-neighborhood of
surface $P$ embedded in $\R^3$. Then obviously $\g=2\T^2$. We may
readily picture a one-parameter family of surfaces $P_{\tau}$, $\tau
\in (-\varepsilon, \varepsilon)$, such that $P_\tau = \T^2$ for
$\tau \ne 0$ and $P_0 = P$. We classify such modification $\T^2 \to
P \to \T^2$ occurring when the parameter $\tau$ changes as type
$(1,1)$. The other modifications are defined similarly, namely
$(2,2)$: $2\T^2 \to Q \to 2\T^2$, $(1,2)$: $\T^2 \to R \to 2\T^2$
and $(0,1)$: $\varnothing \to S^1 \to \T^2$. The last two
modifications, with the order reversed, will be denoted by $(2,1)$
and $(1,0)$. The symbol (1:1) stands for a continuous deformation of
a connected component of $J_n$ on which no critical points of the
integral map arise. Symbols of modifications taking place
simultaneously will be connected by the sign "plus", or will be
supplied by an integer factor if they are of the same type.

We list the sequence of bifurcations occurring along the dashed
arrows in Figures~1 and~2 as follows:

A) 2(0,1), (2,1), (1,2), (1:1) + (1,0), (1,0);

B) (0,1), (1,2);

C) (2,2), 2(1,1);

D) 4(0,1), 2(2,1), (2,1);

E) 2(1,2), 2(1:1) + 2(1,0).

The phase flow on the surfaces $P$ and $R$ is structured as follows:
the middle line of the surface (i.e., the trace of the center of the
"figure-eight") is a periodic solution, while the other trajectories
have this solution as their $\alpha$- and $\omega$-limit cycle. On
the surface $Q$ there are two distinguished closed curves, which are
periodic solutions. The other trajectories approach one of these
solutions asymptotically as $t \to +\infty$, and the other as $t \to
-\infty$. We admit the existence of subset $\sigma \subset \Sigma$
such that all the trajectories on the corresponding surfaces $P$,
$Q$, or $R$ are closed. Due to the non-degeneracy of the problem,
$\sigma$ must have zero measure in $\Sigma$.

\begin{figure}[h]\label{fig3}
\begin{center}
\includegraphics[width=170mm,keepaspectratio]{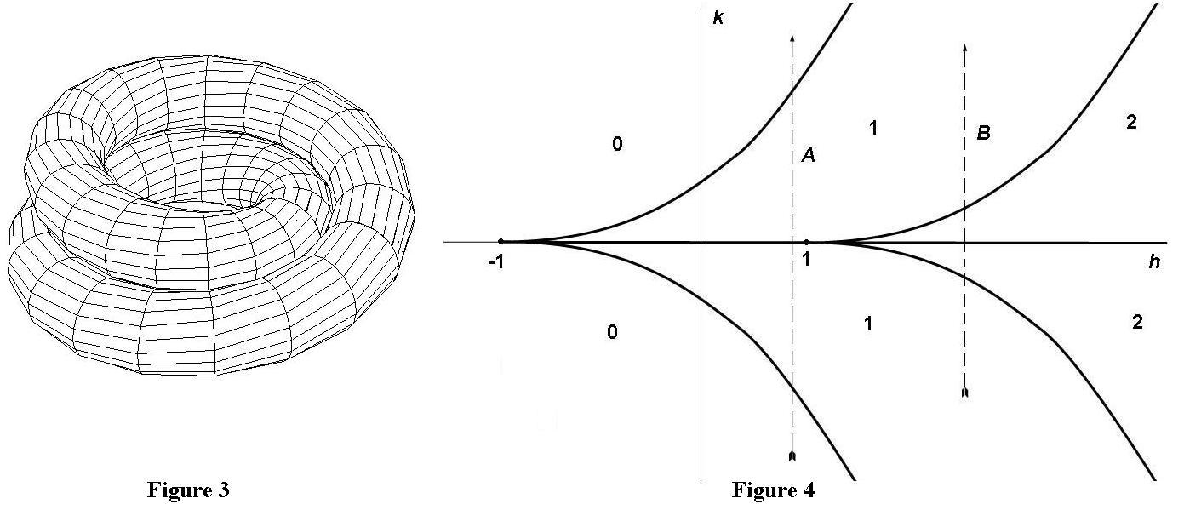}
\end{center}
\end{figure}

2.~In the Goryachev\,--\,Chaplygin case the additional integral is found
under the constraint ${G(\nu, \omega) = 0}$. This equation isolates in
$M$ a four-dimensional manifold $M_0$ diffeomorphic to $T S^2$. In
dimensionless variables, $H$ and $K$ take the form
\begin{equation}\label{eq04}
H = 2(\omega_1^2+\omega_2^2)+\frac{1}{2}\omega_2^2 -\nu_1, \quad K =
\omega_3(\omega_1^2 + \omega_2^2) + \omega_1 \nu_3.
\end{equation}

\begin{theorem}
The bifurcation set $\Sigma_0$ of the map $H \times K|_{M_0}$ is
defined by the equality
$$
\Sigma_0=\{k=0,h\geqslant -1\}\cup\{h=\frac{3}{2}(2k)^{2/3} \pm 1\}.
$$
It divides the $(h, k)$-plane into 5 open components {\rm (}Figure 4{\rm )}. The
corresponding integral manifolds are $\varnothing$, $\T^2$ $($for two
components$)$, and $2\T^2$ {\rm (}for two components{\rm )}.
\end{theorem}

Preserving the notation used above, we write the sequences of
bifurcations occurring along the arrows in Figure 4:\nopagebreak[3]

A) (0,1), (1,1), (1,0);\nopagebreak[3]

B) (1,2), (1:1) + (1,0) + (0,1), (2,1).

Recall that for $k = 0$, besides the first integrals (\ref{eq04}),
there exists another "superfluous" integral, discovered by
D.\,N.\,Goryachev. For this reason the surfaces $P$ appearing on the
segment $\{{k=0,}{-1<h<1}\}$, as well as the regular two-dimensional
torus component of $J_n$ on the segment $\{{k=0,}{h>1}\}$, consist
entirely of periodic trajectories. Moreover, on surface $P$ the
middle line is "twice shorter" than the neighboring periodic
solutions.

\vspace{1cm} \footnotesize

\hbox to 1\textwidth {Volgograd~State~University \hfil
Received~4/MAR/83}

\vspace{2cm}

\normalsize

\renewcommand\refname{BIBLIOGRAPHY}

\begin{flushright}
Translated by A. IACOB

from Doklady Academii Nauk SSSR, vol. 273 (1983), N 6.
\end{flushright}
\end{document}